\documentclass{emulateapj}
\usepackage{apjfonts}

\shorttitle{GOOGS 850-5 --- A Submillimeter Galaxy at $z>4$?}
\shortauthors{Wang et al.}

\begin{document}

\title{GOODS 850-5 --- A $z>4$ Galaxy Discovered in the Submillimeter?}

\author{Wei-Hao Wang,\altaffilmark{1,2}
Lennox L. Cowie\altaffilmark{3},
Jennifer van Saders\altaffilmark{4,2},
Amy J. Barger\altaffilmark{5,6,3},
and Jonathan P. Williams\altaffilmark{3}}

\altaffiltext{1}{Jansky Fellow, {\tt whwang@aoc.nrao.edu}}
\altaffiltext{2}{National Radio Astronomy Observatory,
1003 Lopezville Road, Socorro, NM 87801.  The NRAO
is a facility of the National Science Foundation operated under 
cooperative agreement by Associated Universities, Inc.}
\altaffiltext{3}{Institute for Astronomy, University of Hawaii, 
2680 Woodlawn Drive, Honolulu, HI 96822}
\altaffiltext{4}{Department of Physics and Astronomy, Rutgers, 
the State University of New Jersey,
136 Frelinghuysen Road, Piscataway, NJ 08854-8019}
\altaffiltext{5}{Department of Astronomy, University of Wisconsin-Madison, 
475 North Charter Street, Madison, WI 53706}
\altaffiltext{6}{Department of Physics and Astronomy, 
University of Hawaii, 2505 Correa Road, Honolulu, HI 96822}
\footnotetext[7]{The Submillimeter Array is a joint project between 
the Smithsonian Astrophysical Observatory and the Academia Sinica 
Institute of Astronomy and Astrophysics and is funded by the Smithsonian 
Institution and the Academia Sinica.}
\footnotetext[8]{The NASA/IPAC Extragalactic Database 
(NED) is operated by the Jet Propulsion Laboratory, California Institute of 
Technology, under contract with the National Aeronautics and Space Administration.}
\slugcomment{ApJ Letters accepted, October 2007}

\begin{abstract}
We report an SMA interferometric identification of a bright submillimeter 
source, GOODS 850-5.  This source is one of the brightest 850 $\mu$m sources
in the GOODS-N but is extremely faint at all other wavelengths.  It is
not detected in the GOODS \emph{HST} ACS images and only shows a weak 
2 $\sigma$ signal at 1.4 GHz.  It is detected in the \emph{Spitzer} IRAC 
bands and the MIPS 24 $\mu$m band, however, with very low fluxes.  
We present evidence in the radio, submillimeter, mid-IR,
near-IR, and optical that suggest GOODS 850-5 may be a $z>4$ galaxy.  
\end{abstract}

\keywords{cosmology: observations --- galaxies: evolution --- 
galaxies: formation --- galaxies : starburst --- IR: 
galaxies --- submillimeter}

\section{Introduction}
The Rayleigh-Jeans portion of the dust spectral energy 
distribution (SED) produces a strong negative $K$-correction
and makes the observed submillimeter flux of a dusty galaxy
almost invariant at $z>1$ to $z\sim10$ \citep{blain93}.
This makes the submillimeter wavelength a potentially powerful
probe to the high-redshift universe.  However, to date, all
the identified submillimeter galaxies are at redshifts
lower than 4, likely because of the limited resolution of 
current submillimeter instruments and the limited sensitivity 
of current radio instruments.

The Submillimeter Common-User Bolometer Array (SCUBA) on the 
single-dish James Clerk Maxwell Telescope resolved 20\%--30\% 
of the submillimeter extragalactic background light into point 
sources brighter than $\sim2$ mJy at 850 $\mu$m 
(\citealp{barger99,eales99}).  Because of the low resolution
($\sim15\arcsec$) of SCUBA, identifications of the submillimeter
sources have to assume the radio--FIR correlation in local galaxies 
\citep[see, e.g.,][]{condon92} and rely on radio interferometry
to pinpoint the location of the submillimeter emission. 
Optical spectroscopy of radio identified submillimeter sources
shows that they are ultraluminous ($>10^{12}$ $L_{\sun}$, 
corresponding to star formation rates of $10^2$--$10^3$ 
$M_{\sun}$ yr$^{-1}$) sources at $z\sim2$--3 and that they dominate 
the total star formation at this redshift \citep{chapman05}.  
However, the positive 
$K$-correction of the radio synchrotron emission makes the radio
wavelength insensitive to high-redshift galaxies, and radio 
observations can only identify 60\%--70\% of the blank-field
submillimeter sources \citep[e.g.,][]{barger00,ivison02}.  
The radio unidentified submillimeter sources are commonly thought 
to be at redshifts higher than the radio detection limit 
(typically $z\sim3$--4) but there is no direct evidence 
for such a high-redshift radio-faint submillimeter population.

With recent development in submillimeter interferometry,
it is now possible to locate submillimeter sources directly 
without relying on radio observations.  We have begun a program
to target radio-faint submillimeter sources with the 
Submillimeter Array (SMA\footnotemark[7]) to determine whether 
there is a high-redshift ($z>4$) tail in the redshift distribution 
of the submillimeter sources.  Here we report our first 
identification in this program, GOODS 850-5.  
GOODS 850-5 was detected by our 
SCUBA jiggle-map survey in the Great Observatories Origins 
Deep Surveys-North (GOODS-N, \citealp{giavalisco04a}) with an 
850 $\mu$m flux of $12.9\pm2.1$ mJy \citep{wang04}.
It was also detected in the combined jiggle and scan map
of GOODS-N (GN 10, see \citealp{pope06} and references therein).  
It is the second brightest submillimeter 
source in our jiggle-map catalog of the GOODS-N and has an IR
luminosity of $\sim2\times10^{13}$ $L_{\sun}$.  It does not have a 
$5 \sigma$ radio counterpart in the deep Very Large Array (VLA) 
1.4 GHz catalogs of \citet{richards00} and \citet{biggs06}.  
We obtained an 
unambiguous identification of GOODS 850-5 with the SMA and found 
that the counterpart to this source is remarkably faint at 
\emph{all} other wavelengths.  All data in the optical, near-IR, 
mid-IR, submillimeter, and radio point to a source at $z>4$.  
This is important evidence for the existence 
of high-redshift submillimeter sources.  In this letter, we report 
the SMA observation and the multiwavelength photometry (\S~\ref{observation})  
and the redshift constraints derived 
from the photometric data (\S~\ref{z_constraints}).  We discuss the 
implication for the star formation history in \S~\ref{discussion} and 
summarize in \S~\ref{summary}.  We adopt $H_0=71$ km s$^{-1}$ Mpc$^{-1}$,
$\Omega_{\Lambda}=0.73$, and $\Omega_M=0.27$.

\begin{figure*}
\epsscale{1.17}
\plotone{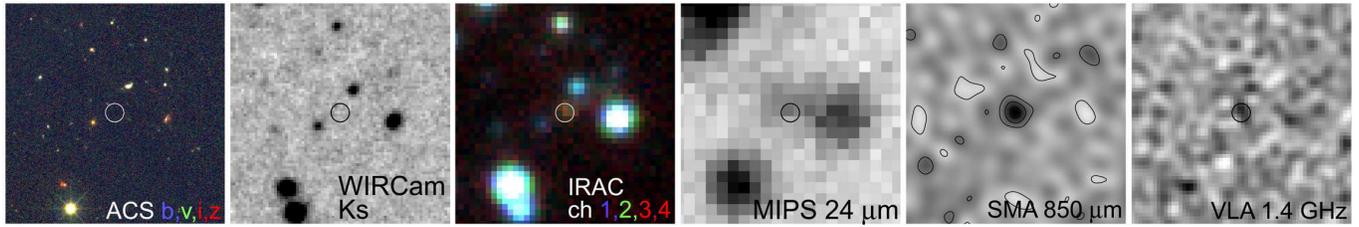}
\caption{Multi-wavelength images of GOODS 850-5.  Each panel has
a size of $24\arcsec$.  North is up.  All four of the ACS
(F435W, $b$; F606W, $v$; F775W, $i$; and F850LP, $z$) bands and 
all four of the IRAC channels (1, 2, 3, and 4) are included in
the color pictures.  The color codes are labeled in the images.
Grayscale images have inverse scales.  The SMA position is 
labeled with $2\arcsec$ diameter circles.  The contours in the 
SMA image have levels of -2, 2, 4, and 6 $\sigma$ where 1 $\sigma$ 
is 1.4 mJy beam$^{-1}$. \label{thumbnail}}  
\end{figure*}

\section{Observations and Measurements}\label{observation}
A full track of SMA observations were obtained on 2007 January 23 
using eight antennas in the compact configuration.  The receivers 
were tuned to a frequency of 345.0 GHz centered in the upper 
sideband, yielding 335.0 GHz in the lower sideband.  Each sideband 
covers a 2 GHz bandwidth and contains 768 spectral channels.  
The tracking center is at the SCUBA position of GOODS 850-5 
($\alpha\rm[J2000.0] =\rm 12^h 36^m 33^s.45$, 
$\delta\rm[J2000.0]=62\arcdeg 14\arcmin 09\farcs43$).
Callisto and the bright quasar 3C84 were observed to provide 
flux and bandpass calibrations, respectively.  Quasars 
1419+543 and 1048+717, respectively $15.\mkern-6mu\arcdeg5$ and 
$14\arcdeg$ away from the target, were observed after every 15 minutes 
of on-target integration for time-dependent complex gain calibrations.
The 225 GHz opacity was an excellent 0.03--0.06, and the dual-sideband
system temperature averaged over the entire track was 188 K.

The calibration and data inspection were performed with the Caltech 
package MIR, modified for the SMA.  Continuum data were generated
by averaging the spectral channels after the passband calibration.
Both gain calibrators were used to derive gain curves.  However,
we carried out a separate reduction in which the primary
calibrator 1048+717 was used to calibrate 1419+543.  The reduced
image of 1419+543 has a point source $0\farcs24$ away from 
the phase center.  Since the two calibrators are separated 
by a large $28\arcdeg$ on the sky, this offset should be only
considered as an upper limit to the astrometry error in the 
reduced image of GOODS 850-5.
The calibrated visibility data were Fourier transformed and
deconvolved in the package MIRIAD to form images.  In the 
transformation, we weighted each visibility point inversely 
proportional to the system temperature.  Combined with a natural 
weighting, this provides an optimum S/N and a synthesized beam 
FWHM of $2\farcs4 \times 2\farcs2$ at a position angle of $24\arcdeg$.
In the dirty image of GOODS 850-5, a point source is apparent and 
we restricted the CLEAN deconvolution to a $\sim10\arcsec$ region 
centered at the point source.  There is no sign of other sources 
both in the dirty and cleaned images.  The noise level in the
cleaned image is 1.4 mJy beam$^{-1}$.

A point source of $12.0\pm1.4$ mJy is clearly detected very close
to the phase center.  A Gaussian fit to the image yields a
source profile identical to the synthesized beam, indicating
an unresolved source with good calibrations.  The fit provides
a source position of $\alpha(\rm J2000.0) = 12^h 36^m 33^s.45$, 
$\delta(\rm J2000.0) = 62\arcdeg 14\arcmin 08\farcs65$, which 
agrees extremely well with the SCUBA position.  The positional 
error in the fit is $0\farcs2$, consistent with the 
angular resolution for this S/N and also comparable to the
aforementioned upper limit to the astrometry error derived from the 
calibrator 1419+543.  We also performed a point-source fit directly 
to the visibility data, and the result is fully consistent with 
the fit in the image plane.  

We present the GOODS \emph{HST} Advanced Camera for Surveys (ACS) 
images \citep{giavalisco04a}, our CFHT WIRCam $K_s$ band image 
(36 hours of integration, Keenan et al.\ 2007, in preparation),
the GOODS \emph{Spitzer} Legacy Program 
(Dickinson et al.\ 2007, in preparation) Infrared Array Camera
(IRAC) and Multiband Imaging Photometer for \emph{Spitzer} (MIPS) 
images, the VLA image \citep{biggs06}, and our SMA image of 
GOODS 850-5 in Fig.~\ref{thumbnail}.  In all of the four ACS 
images (F435W, F606W, F775W, and F850LP) and the WIRCam image, 
there are no galaxies within $0\farcs5$ from the SMA position.  
A faint galaxy is clearly detected by \emph{Spitzer} in all four 
IRAC bands and in the 24 $\mu$m MIPS band right at the SMA position.  
Because of the high accuracy ($0\farcs2$) of the SMA astrometry, 
this \emph{Spitzer} source is a robust identification of GOODS 850-5.
We list our broadband fluxes in Table~\ref{tab1}.  The
850 $\mu$m flux is a noise weighted mean from the SCUBA jiggle map
(Wang et al.\ 2004) and the SMA measurement.  Typical errors in
flux calibrations at this wavelength are $\gtrsim10\%$ and could be
larger than the noise.  However, the fact that the SMA flux and
the SCUBA flux agree with each other extremely well suggests good 
flux calibrations in both experiments.  The 1.4 GHz flux is measured 
at the SMA position from the VLA image of \citet{biggs06}, 
assuming a point source and with a primary beam correction.
The ACS and WIRCam fluxes are also measured at the SMA position
with respectively $0\farcs8$ and $1\farcs5$ diameter apertures.
Measuring fluxes in the \emph{Spitzer} bands is more difficult.
In the 3.6 and 4.5 $\mu$m images, there is a nearby blue
galaxy blended with GOODS 850-5.  There is another galaxy
blended with GOODS 850-5 in the MIPS image.  To deconvolve the fluxes, 
we performed PSF fitting to the blended sources.  We used the
IRAC and MIPS in-flight PSFs and applied simultaneous 
two-component fits to the two sources, and scaled the measured
fluxes to total fluxes using the PSFs.  To verify the fluxes
measured in this way, we first compared them with simple 
aperture-corrected aperture fluxes.  We found that at 5.8
and 8.0 $\mu$m where GOODS 850-5 is isolated, the PSF fitting 
provides fluxes consistent with the aperture fluxes.  
We also used the SExtractor package \citep{bertin96} to measure fluxes.  
SExtractor could deblend the two sources at 4.5 $\mu$m but not
at 3.6 $\mu$m even with the most aggressive deblending.
The aperture corrected SExtractor fluxes at $4.5$ $\mu$m and 
longer are all consistent with the PSF fitting fluxes.  
We thus conclude that our PSF fitting provides reliable estimates
of the fluxes of GOODS 850-5.  The \emph{Spitzer} fluxes listed 
in Table~\ref{tab1} are all measured with the PSF fitting method.  

We note that our SMA identification confirms the counterpart 
suggested by \citet{pope06}.  Our carefully measured \emph{Spitzer} 
fluxes are also consistent with the SExtractor pipeline fluxes in 
Pope et al.  The 1.4 GHz flux in Pope et al.\ is 76\% larger than
ours but the difference is within the errors.

\begin{deluxetable*}{cccccccccccc}[ht!]
\tablecaption{Photometry of GOODS 850-5\label{tab1}}
\tabletypesize{\small}
\tablewidth{0pt}
\tablehead{
\colhead{\scriptsize $S_{\rm F435W}$} &
\colhead{\scriptsize $S_{\rm F606W}$} &
\colhead{\scriptsize $S_{\rm F775W}$} &
\colhead{\scriptsize $S_{\rm F850LP}$} &
\colhead{\scriptsize $S_{Ks}$} &
\colhead{\scriptsize $S_{3.6\mu\rm m}$} &
\colhead{\scriptsize $S_{4.5\mu\rm m}$} &
\colhead{\scriptsize $S_{5.8\mu\rm m}$} &
\colhead{\scriptsize $S_{8.0\mu\rm m}$} &
\colhead{\scriptsize $S_{24\mu\rm m}$} &
\colhead{\scriptsize $S_{850\mu\rm m}$} &
\colhead{\scriptsize $S_{1.4\rm GHz}$} }
\startdata
\scriptsize -0.013$\pm$0.004 & \scriptsize -0.004$\pm$0.003 
& \scriptsize 0.001$\pm$0.006 & \scriptsize -0.009$\pm$0.009 & 
 \scriptsize 0.055$\pm$0.052 & \scriptsize 1.14$\pm$0.14 & \scriptsize 1.64$\pm$0.13 
& \scriptsize 2.33$\pm$0.24 & \scriptsize 5.37$\pm$0.37 &  \scriptsize
46.3$\pm$9.2 & \scriptsize 12300$\pm$1200 & \scriptsize 18.7$\pm$8.0
\enddata
\tablecomments{All fluxes are in $\mu$Jy, which can be converted
to AB magnitude with $AB=23.9-2.5\log(\mu \rm Jy)$.}
\end{deluxetable*}

\section{Redshift Constraints}\label{z_constraints}

GOODS 850-5, one of the brightest 850 $\mu$m sources in the GOODS-N,
is incredibly faint at \emph{all} other wavelengths.  Because of the
optical and near-IR faintness, it is difficult to obtain
its redshift through spectroscopy.  We obtained redshift constraints
for GOODS 850-5 using the multiwavelength broadband fluxes.  There are
three independent pieces of evidence that suggest GOODS 850-5 is a
$z>4$ galaxy.  

By assuming that the well known radio--FIR correlation for local
star forming galaxies also applies to submillimeter sources, we
can estimate the redshift using the radio and submillimeter fluxes
(e.g., \citealp{carilli99}; Barger et al.\ 2000).
Barger et al.\ assumed pure power-law SEDs in both the radio and
the submillimeter and obtained the formula 
$z=0.98(S_{850 \mu \rm m}/S_{1.4 \rm GHz})^{0.26} - 1$, based on
Arp 220.  This implies $z\sim4.3$ for GOODS 850-5.  This is a lower 
limit, because the peak of the dust SED and the free-free emission 
start to move in the observed bands at $z>4$.  Both effects
underestimate the redshift.  By adopting the complete Arp 220 
SED model in \citet{silva98}, we find $z\sim5.6$ for 
GOODS 850-5 ($z\sim4.3$ if we adopt the radio flux of 33.1 $\mu$Jy 
in \citealt{pope06}).  This redshift estimate is nevertheless very 
crude.  Because of the uncertainties in the submillimeter and radio SEDs
and the radio--FIR correlation at high redshift, such redshift 
estimates typically have an error of $\pm0.5$ \citep[e.g.,][]{ivison02}.  


\begin{figure}[h!]
\epsscale{0.8}
\plotone{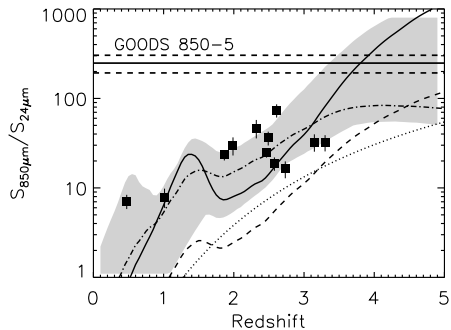}
\caption{$S_{850\mu \rm m}/S_{24\mu \rm m}$ flux ratio vs.\ redshift.  
Curves are filter-convolved \citet{silva98} model SEDs of Arp 220 
(ultraluminous starburst, \emph{solid curve}), M 100 (normal spiral,
\emph{dash-dotted curve}), and M 82 (low luminosity starburst, 
\emph{dashed curve}), and Mrk 231 (ultraluminous dusty AGN, 
\emph{dotted curve}) derived from the photometry in NED$^8$.  
Shaded region is the range of the models in \citet{chary01}.  
\emph{Solid squares} are GOODS-N SCUBA sources identified at
24 $\mu$m or 1.4 GHz \citep{wang06}.  \emph{Horizontal lines} are 
GOODS 850-5 and the 1 $\sigma$ error.  \label{850_24_z}}  
\end{figure}

The unusually large $S_{850 \mu \rm m}/S_{24 \mu \rm m}$ ratio 
also suggests a high redshift of $z>3$.  As can be seen in 
Fig.~\ref{850_24_z}, the $S_{850 \mu \rm m}/S_{24 \mu \rm m}$ 
ratio of GOODS 850-5 is an order of magnitude higher than 
the mean value for other identified submillimeter sources 
in the GOODS-N.  This can only be explained by a $z>3$ galaxy 
with any dust SEDs allowed by current models.

\begin{figure}[h!]
\epsscale{0.95}
\plotone{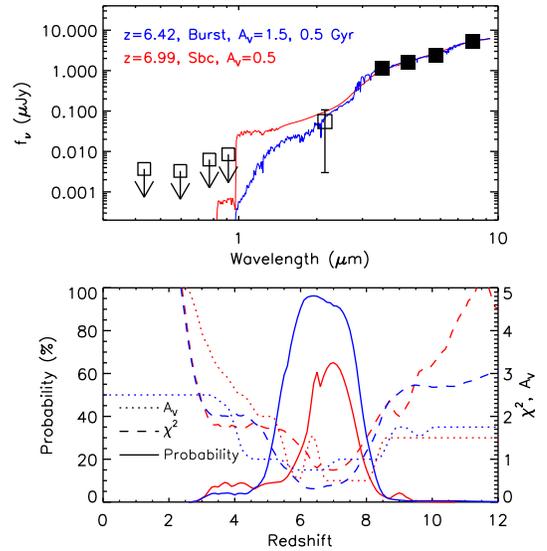}
\caption{Photometric redshift result for GOODS 850-5.  
The \emph{upper panel} shows the observed IRAC, WIRCam, and ACS SEDs 
(\emph{squares}), and the best-fit SEDs (\emph{solid curves}) 
in the empirical set (\emph{red}) and in the population synthesis 
set (\emph{blue}).  
The \emph{lower panel} shows the $\chi^2$ (\emph{dashed curves}), 
the corresponding probability (\emph{solid curves}), and the 
extinction (\emph{dotted curves}) of the best-fit SEDs vs.\ 
redshift for the two SED sets (same color coding as in
the upper panel).  \label{zphot}}  
\end{figure}

Finally, we used the photometric redshift technique to test if
the \emph{HST}, WIRCam, and \emph{Spitzer} SED of GOODS 850-5 
can be fitted with a $z>4$ galaxy.
Because there is not a unique correlation between the dust 
SED and stellar SED of galaxies, we only included the ACS, WIRCam,
and IRAC passbands in the photometric redshift study to account
for just the stellar emission.  Although 
GOODS 850-5 is not detected in the ACS images, the upper limits to 
the ACS fluxes suggest a very red SED at $<3$ $\mu$m 
(Fig.~\ref{zphot}) and thus still provide useful constraints.
In the photometric redshift fitting, we assigned zeros to the ACS 
fluxes and used the 1 $\sigma$ errors as the flux uncertainties.
We used the measured $K_s$ flux, even though it is only 1 $\sigma$.

We used the popular Hyperz package \citep{bolzonella00}.  
We limited the redshift range to $z=0$--12 and adopted the 
\citet{calzetti00} 
extinction of $A_V=0$--5.  We used two sets of SEDs.  The first set 
consists of empirically observed SEDs, including the Im, Scd, Sbc, and E 
type galaxies in Coleman et al.\ (1980) and the starburst SEDs SB2 
and SB3 in \citet{kinney96}.  The primary solution provided by Hyperz 
is $z=6.99$, $A_V=0.5$, and a Sbc type galaxy.  The corresponding 
probability is 65.1\%, and the 99\% confidence range is $z=5.6$--8.0.  
The second SED set is the stellar population synthesis models 
included in Hyperz, which is based on the earlier 
version of \citet{bc93}.  The primary solution is 
$z=6.42$, $A_V=1.5$, and a burst type SED with an age of 0.5 Gyr.  
The corresponding probability is 96.4\%, and the 99\% confidence
range is $z=5.2$--8.0.  In both SED sets, 
the photometric redshift confidence range is consistent with the 
range allowed by the radio, submillimeter, and mid-IR 
flux ratios discussed above.  In Fig.~\ref{zphot} we show the fitted 
and observed SEDs, and the best-fit $\chi^2$, probability, and 
extinction as functions of redshift.  The SED can only be fitted 
with $A_V>2$ at $z<3$ and the probabilities at $z<3$ of the best-fit 
models are zero even when $A_V=10$ is allowed in the fitting.  
Therefore, a highly reddened galaxy at $z<3$ is ruled out for 
GOODS 850-5.  In the photometric redshift fitting, an intermediate 
redshift of 3--4.5 is not entirely ruled out, although the probability 
is low ($<10\%$).

The primary uncertainty in the above analyses is the possibility of 
the existence of an active galactic nucleus (AGN), whose SED may
be different then the ones used for the radio, submillimeter,
IR, and optical photometric analyses.  We argue that
whether or not there is an AGN is not likely to change the high 
redshift of GOODS 850-5.  First, GOODS 850-5 is not detected 
in the 2 Ms \emph{Chandra} images \citep{alexander03} in 
either the hard or soft X-ray bands.  This rules out a 
Compton-thin AGN at $z<3$.  (The 2 Ms \emph{Chandra} observations 
can detect a $10^{42}$ and $10^{42.5}$ erg s$^{-1}$ soft and hard 
X-ray source, respectively, at $z\sim3$.)  The best example of 
a local Compton-thick AGN in the far-IR luminosity class of 
GOODS 850-5 is Mrk 231.  Although the IRAC power-law
spectral slope of GOODS 850-5 is similar to that of Mrk 231, 
its $S_{850\mu \rm m}/S_{24\mu \rm m}$ flux ratio
is more than an order of magnitude higher than Mrk 231 at $z<3$
(Fig.~\ref{850_24_z}), i.e., GOODS 850-5 does not show much of 
the warm dust emission that is characteristic for a
heavily obscured AGN.  Furthermore, the rest-frame 1.4 GHz 
flux of Mrk 231 also follows the radio--FIR correlation for 
normal galaxies with a $q$ value of 2.24 (under the definition of 
$q$ in \citealt{condon92}) but it has a flat radio spectrum
(a smaller $K$-correction in the radio).  Therefore, even if 
GOODS 850-5 is an AGN that is similar to Mrk 231, its 
large $S_{850 \mu \rm m}/S_{1.4 \rm GHz}$ flux ratio will 
suggest a redshift even higher than that estimated from the
Arp 220 template.

\section{Implications for the Star Formation History}\label{discussion}

There is growing evidence for a signification submillimeter population 
at $z>4$.  \citet{dunlop04} presented evidence for an optically
faint $z\sim4.1$ galaxy for HDF 850.1.  More recent SMA results of 
\citet{younger07} suggest a number of radio and optically 
faint submillimeter sources to be at $z\gtrsim4$.  GOODS 850-5 is 
similar to these galaxies.  On the other hand, GOODS 850-5 has the 
highest $S_{850\mu\rm m}/S_{1.4\rm GHz}$ flux ratio, and also has higher 
quality data at many other wavelengths.  Unlike HDF 850.1, which
is behind a bright galaxy, GOODS 850-5 can be studied in details
especially in the optical and IR wavelengths.  

The discovery of GOODS 850-5 and its like extends the redshift 
distribution of bright submillimeter sources to beyond 4.0, and perhaps 
even to $z>5$.  The 12.3 mJy 850 $\mu$m flux of GOODS 850-5 
implies an IR luminosity of $\sim2.3\times10^{13}$ $L_{\sun}$ 
($L_{\rm IR} = 1.9 \times 10^{12} S_{850 \mu \rm m} L_{\sun}/$mJy,
\citealp{blain02}), and therefore a very high star formation rate 
of $\sim4000$ $M_{\sun}$ yr$^{-1}$
($\dot{M}=1.7\times10^{-10} L_{\rm IR}/L_{\sun}$, \citealp{kennicutt98}).
This is an extraordinary amount of star formation at high redshift
but it can only be identified in the submillimeter.  Even if such a 
source could be picked up in optical and near-IR broadband 
dropout surveys, its star formation would still be missed due to 
the extreme faintness in the rest-frame UV.
How much can hidden sources like this affect the current picture of
the cosmic star formation history at high redshift?  

Bright submillimeter sources ($>2$ mJy at 850 $\mu$m) 
contribute $\sim10$ Jy deg$^{-2}$ to the total background, 
and 60\% to 70\% of them have radio counterparts.  We assume 
that all of the $\sim4$ Jy deg$^{-2}$ radio-unidentified bright 
submillimeter sources are at $z=4$--7.  With the above conversions 
between 850 $\mu$m flux, IR luminosity, and star formation rate, 
this implies a mean comoving IR luminosity density of 
$\sim2.2\times10^8$ $L_{\sun}$ Mpc$^{-3}$, and therefore a star 
formation rate density (SFRD) of $\sim0.04$
$M_{\sun}$ yr$^{-1}$ Mpc$^{-3}$.  This is consistent with the 
estimate in Barger et al.\ (2000)
based on a smaller sample of radio unidentified submillimeter sources, 
after taking into account the differences in cosmology.
Our value of 0.04 $M_{\sun}$ yr$^{-1}$ Mpc$^{-3}$ \emph{averaged over}
$z=4$--7 is $\sim2\times$ higher than the typical value 
of 0.01--0.02 $M_{\sun}$ yr$^{-1}$ Mpc$^{-3}$ \emph{at} $z\sim4$ 
determined from optical surveys without extinction correction 
\citep[e.g.,][]{giavalisco04b,bouwens03}.  
In general, Lyman-break galaxies are not bright submillimeter 
sources \citep{peacock00,chapman00,webb03}, suggesting that most  
(if not all) of the above SFRD estimated from the radio-unidentified 
bright submillimeter sources is missed by optical surveys.
Although our SFRD of 0.04 $M_{\sun}$ yr$^{-1}$ Mpc$^{-3}$ is likely an 
upper limit due to the assumptions, it still shows the significance of 
dust hidden star formation at high redshift.  

\section{Summary}\label{summary}
We obtained accurate astrometry for the bright submillimeter
source GOODS 850-5 with the SMA interferometer at 850 $\mu$m.  
The counterpart is extremely faint in the optical, near-IR, mid-IR, 
and radio.  Both the radio and 24 $\mu$m faintness of GOODS 850-5 
suggests a high redshift of $z>4$.  The very red SED between
3.6 and 8.0 $\mu$m and the nondetection in the optical and $K_s$ bands
can only be fitted by stellar continuum at $z>4$.  This discovery
provides important evidence that some of the radio 
undetected submillimeter sources are at high redshift and extends 
the redshift distribution of bright submillimeter sources to $z>4$.
It suggests that a great fraction of star formation at high
redshift is hidden from optical surveys.

\acknowledgments
We thank L. Silva and R. Chary for providing the SED templates,
the SMA staffs for the help on observation and data reduction,
and the referee for the comments that greatly improve the manuscript.
We gratefully acknowledge support from NRAO (W.-H.W.), 
NRAO REU program (J.V.S.), NSF grants AST 04-07374 (L.L.C.) and 
AST 02-39425 (A.J.B.), the University of Wisconsin Research Committee 
with funds granted by the Wisconsin Alumni Research Foundation, 
and the David and Lucile Packard Foundation (A.J.B.).


\begin{thebibliography}{} 

\bibitem[Alexander et al.(2003)]{alexander03} 
   Alexander, D.\ M.\ et al.\ 2003, \aj, 126, 539
\bibitem[Barger, Cowie, \& Sanders(1999)]{barger99}
   Barger, A.\ J., Cowie, L.\ L., \& Sanders, D.\ B.\ 1999, \apjl, 518, L5
\bibitem[Barger, Cowie, \& Richards(2000)]{barger00}
   Barger, A.\ J., Cowie, L.\ L., \& Richards, E.\ A.\ 2000, \aj, 119, 2092
\bibitem[Bertin \& Arnouts(1996)]{bertin96} 
   Bertin, E.\ \& Arnouts, S.\ 1996, \aaps, 117, 393
\bibitem[Biggs \& Ivison(2006)]{biggs06}
   Biggs, A.\ D.\ \& Ivison, R.\ J.\ 2006, \mnras, 371, 963
\bibitem[Blain \& Longair(1993)]{blain93}
   Blain, A.\ W.\ \& Longair, M. S. 1993, \mnras, 264, 509
\bibitem[Blain et al.(2002)]{blain02}
   Blain, A.\ W., Smail, I., Ivison, R.\ J., Kneib, J.-P., \&
   Frayer, D.\ T.\ 2002, \physrep, 369, 111
\bibitem[Bolzonella, Miralles, \& Pell\'{o}(2000)]{bolzonella00}
   Bolzonella, M., Miralles, J.-M., \& Pell\'{o}, R. 2000, \aap, 363, 476
\bibitem[Bouwens, Broadhurst, \& Illingworth(2003)]{bouwens03}
   Bouwens, R., Broadhurst, T., \& Illingworth, G. 2003, \apj, 593, 640
\bibitem[Bruzual \& Charlot(1993)]{bc93}
   Bruzual, G., \& Charlot, S.\ 1993, \apj, 405,538
\bibitem[Calzetti et al.(2000)]{calzetti00}
   Calzetti, D., Armus, L., Bohlin, R.\ C., Kinney, A.\ L.,
   Koornneef, J., \& Storchi-Bergmann, T.\ 2000, \apj, 533, 682
\bibitem[Carilli \& Yun(1999)]{carilli99}
   Carilli, C.\ L.\ \& Yun, M.\ S.\ 1999, \apjl, 513, L13
\bibitem[Chapman et al.(2000)]{chapman00} Chapman, S.\ C., et al.\ 2000, \mnras, 319, 318
\bibitem[Chapman et al.(2005)]{chapman05}
   Chapman, S.\ C., Blain, A.\ W., Smail, I., \& Ivison, R.\ J.\ 2005, \apj, 
   622, 772
\bibitem[Chary \& Elbaz(2001)]{chary01} Chary, R.\ \& Elbaz, D.\ 2001, \apj, 556, 562
\bibitem[Coleman, Wu, \& Weedman(1980)]{cww80}
   Coleman, G.\ D., Wu, C.-C., \& Weedman, D.\ W.\ 1980, \apjs, 43, 393
\bibitem[Condon(1992)]{condon92} Condon, J.\ J.\ 1992, \araa, 30, 575
\bibitem[Dunlop et al.(2004)]{dunlop04}
   Dunlop, J.\ S., et al.\ 2004, \mnras, 350, 769
\bibitem[Eales et al.(1999)]{eales99}
   Eales, S., Lilly, S., Gear, W., Dunne, L., Bond, J.\ R., Hammer, F., 
   F\`{e}vre, O.\ L., \& Crampton, D.\ 1999, \apj, 515, 518
\bibitem[Giavalisco et al.(2004a)]{giavalisco04a} 
   Giavalisco, M.\ et al.\ 2004a, \apjl, 600, L93
\bibitem[Giavalisco et al.(2004b)]{giavalisco04b} 
   Giavalisco, M.\ et al.\ 2004b, \apjl, 600, L103
\bibitem[Ivison et al.(2002)]{ivison02} Ivison, R.\ J., et al.\ 2002, \mnras, 337, 1
\bibitem[Kennicutt(1998)]{kennicutt98} Kennicutt, R. C. 1998, \araa, 36, 189
\bibitem[Kinney et al.(1996)]{kinney96}
   Kinney, A.\ L., Calzetti, D., Bohlin, R.\ C., McQuade, K., 
   Storchi-Bergmann, T., \& Schmitt, H.\ R.\ 1996, \apj, 467, 38
\bibitem[Peacock et al.(2000)]{peacock00}
   Peacock, J.\ A., et al.\ 2000, \mnras, 318, 535
\bibitem[Pope et al.(2006)]{pope06} Pope, A., et al.\ 2006, \mnras, 370, 1185
\bibitem[Richards(2000)]{richards00} Richards, E.\ A.\ 2000, \apj, 533, 611
\bibitem[Silva et al.(1998)]{silva98}
   Silva, L., Granato, G.\ L., Bressan, A., \& Danese, L.\ 1998, \apj, 509, 103
\bibitem[Webb et al.(2003)]{webb03} Webb, T.\ M., et al.\ 2003, \apj, 582, 6
\bibitem[Wang, Cowie, \& Barger(2004)]{wang04}
   Wang, W.-H., Cowie, L.\ L.\ \& Barger, A.\ J.\ 2004, \apj, 613, 655
\bibitem[Wang, Cowie, \& Barger(2006)]{wang06}
   Wang, W.-H., Cowie, L.\ L.\ \& Barger, A.\ J.\ 2006, \apj, 647, 74
\bibitem[Younger et al.(2007)]{younger07}
   Younger, J.\ D., et al.\ 2007, \apj, in press (arXiv:0708.1020)

\end{thebibliography}
\end{document}